\documentclass[12pt,letterpaper]{article}
\usepackage{amssymb}
\usepackage{amsfonts}
\usepackage{amsmath}
\usepackage{amsthm}
\usepackage[letterpaper,nohead]{geometry}
\usepackage{setspace}
\usepackage[bottom]{footmisc}
\usepackage{indentfirst}
\usepackage{graphicx}
\usepackage{microtype}
\usepackage{needspace}
\usepackage[section]{placeins}
\usepackage{rotating}
\usepackage{enumerate}
\usepackage{booktabs}
\usepackage[flushleft]{threeparttable}
\usepackage{float}
\usepackage{verbatim}
\usepackage{epstopdf}
\usepackage{pdflscape}
\usepackage{mathtools}
\usepackage{mathrsfs}

\newtheorem{theorem}{Theorem}[section]

\newtheorem{proposition}{Proposition}
\newtheorem{corollary}{Corollary}

\newtheorem{remark}{Remark}

\usepackage[usenames,dvipsnames]{color}
\usepackage[round]{natbib}
\usepackage[hyperfootnotes=false]{hyperref}
\hypersetup{
  colorlinks,
  citecolor=Blue,
  linkcolor=Blue,
  urlcolor=Blue}

\newcommand{\Perp}{\perp \! \! \! \perp}

\numberwithin{equation}{section}

\providecommand{\JBESblind}{0}
\begin{document}
\pagenumbering{roman} % Separate title-page anchor from text page 1.
\begin{spacing}{1.2}
\ifnum\JBESblind=1\relax
  \hypersetup{pdftitle={Robust Tests for Heteroskedasticity in Nonparametric Regression under Covariate-Dependent Higher-Order Moments},pdfauthor={}}
  \begin{center}
  {\LARGE\bfseries Robust Tests for Heteroskedasticity in Nonparametric Regression under Covariate-Dependent Higher-Order Moments}
  \end{center}
\else
  \hypersetup{pdftitle={Robust Tests for Heteroskedasticity in Nonparametric Regression under Covariate-Dependent Higher-Order Moments},pdfauthor={Yongzhen Feng and Xiaojun Song}}
  \title{\bfseries Robust Tests for Heteroskedasticity in Nonparametric Regression under Covariate-Dependent Higher-Order Moments}
  \date{}
\author{Yongzhen Feng\thanks{School of Mathematics and Statistics, Beijing Technology and Business University, Beijing, China. E-mail: \texttt{20250707@btbu.edu.cn}. The author acknowledges financial support from the National Natural Science Foundation of China (Grant No. 12601493).}\\Beijing Technology and Business University
\and Xiaojun Song\thanks{Corresponding author. Department of Business Statistics and Econometrics, Guanghua School of Management, Peking University, Beijing, 100871, China. Email: \texttt{sxj@gsm.pku.edu.cn}. This work was supported by the National Natural Science Foundation of China [Grant Numbers 72373007, 72333001, and 72621002]. The author also gratefully acknowledges the research support from the Center for Statistical Science of Peking University and the Key Laboratory of Mathematical Economics and Quantitative Finance (Peking University) of the Ministry of Education, China.}\\Peking University}
\maketitle
\fi

\begin{abstract}
We propose Cram\'er--von Mises- and Kolmogorov--Smirnov-type tests of constant conditional variance in nonparametric regression. The test statistics are constructed from a third-order $U$-process and remain valid when higher-order features of the conditional error distribution vary with the covariates. We derive their limiting distributions under the null hypothesis and study their asymptotic behavior under fixed alternatives and Pitman-type local alternatives. A multiplier bootstrap procedure calibrates the critical values, and Monte Carlo experiments show that the proposed tests have rejection frequencies close to the nominal level in the simulation designs. An application to the National Child Development Study provides evidence of heteroskedasticity in pooled weekly earnings. In the male hourly-earnings sample, rejecting a stronger error-distribution restriction does not entail rejecting homoskedasticity.
\end{abstract}
\end{spacing}

\noindent{\small\emph{Preprint version note.} This version contains the main article only. The supplementary material with detailed proofs and additional simulation and empirical analyses is not included.}



\thispagestyle{empty}

\newpage
\pagenumbering{arabic} \setcounter{page}{1} \setcounter{footnote}{0}

\normalsize

\section{Introduction}\label{sec:intro}

Testing for heteroskedasticity is a standard component of specification analysis in nonparametric regression. The assumption of constant conditional variance matters for conventional inference based on homoskedasticity, while information about the variance function can be exploited to improve estimation efficiency \citep{Glejser1969,Cook1983,Eicker1963,White1980,Robinson1987}. The hypothesis of conditional homoskedasticity concerns only the conditional second moment of the regression error and does not require its entire conditional distribution to remain unchanged across covariate values. In particular, conditional skewness, kurtosis, and other distributional features may vary even when the conditional variance is constant. Consequently, a test of invariance of the entire error distribution may reject even under homoskedasticity. This distinction motivates testing procedures that directly assess the constancy of conditional variance without imposing stronger restrictions on the conditional error distribution.

Consider the nonparametric location-scale regression model
\begin{equation}
Y=m(X)+\varepsilon,\qquad \varepsilon=\sigma(X)\eta,\label{lc}
\end{equation}
where $m(\cdot)$ is the conditional mean function, $\sigma(\cdot)$ is a positive scale function, and $\eta$ is independent of $X$ with mean zero and variance one. In this model, $\operatorname{Var}(Y\mid X)=\sigma^2(X)$, so the hypothesis of homoskedasticity can be expressed as
\begin{equation*}
H_0:\ \sigma(X)\equiv\sigma_0\quad \text{a.s. for some constant }\sigma_0>0.
\end{equation*}
Since the standardized error $\eta$ is independent of the covariates, homoskedasticity in the location-scale model \eqref{lc} also implies independence between $\varepsilon$ and $X$. This motivates residual-based methods that exploit restrictions on error distributions. For example, \citet{EVK2008} develop specification tests for the location-scale model specified in \eqref{lc}, while \citet{Chown2018,Chown2019} use a weighted residual empirical process to test for heteroskedasticity in the above location-scale framework. The latter test does not require a parametric form for the scale function, but its validity is established under the location-scale representation, and the correction in \citet{Chown2019} shows that its limit is not distribution free.

In general nonparametric regression, testing homoskedasticity is not equivalent to testing independence between the regression error and the covariates. Our analysis of earnings and height using data from the National Child Development Study (NCDS) provides a practical illustration of this distinction. For male hourly earnings, the fixed-scale residual-independence test of \citet{EVK2008} rejects the restriction of an invariant error distribution, whereas our tests of constant conditional variance do not reject for the same sample. This contrast illustrates that rejection of a stronger distributional restriction does not by itself constitute evidence against homoskedasticity. Furthermore, our analysis provides evidence against homoskedasticity in pooled weekly earnings, although the tests do not reject in either sex-specific subsample.

Motivated by these considerations, we formulate the testing problem under the general nonparametric regression model
\begin{equation*}
Y=m(X)+\varepsilon,
\end{equation*}
where $m(X)=\mathbb E(Y\mid X)$ and $\mathbb E(\varepsilon\mid X)=0$ almost surely. The regression error is not required to be independent of the covariates in this general framework. The null hypothesis is that the conditional variance of $\varepsilon$ is constant across covariate values, without imposing constancy of its higher-order conditional moments. Our objective is to develop tests of the conditional second-moment restriction that remain valid when these higher-order features depend on the covariates.

The literature has approached the problem of testing conditional variance restrictions from several methodological perspectives, including $L^2$-distance comparisons, nonparametric smoothing, and regression-based specification diagnostics. \citet{Dette1998} study $L^2$ departures from constant variance, \citet{Dette2002} develops a kernel-based test, and \citet{Wooldridge1990} provides a regression-based approach to specification testing. More recently, \citet{KotekalKundu2025} investigate optimal separation rates for heteroskedasticity testing. Nevertheless, several gaps remain for the present setting. The $L^2$-based test of \citet{Dette1998} for fixed designs can detect local alternatives only at the $n^{-1/4}$ rate (see also \citealp{DetteHetzler2009}), whereas kernel-based procedures may be sensitive to bandwidth selection. Moreover, results developed for particular model classes or design settings do not automatically extend to general nonparametric regression with random covariates.

Against this background, three closely related studies are particularly relevant because they illustrate the roles of distributional assumptions, local-alternative sensitivity, and critical-value calibration in heteroskedasticity testing. \citet{Dette2007} compare empirical distributions of standardized residuals obtained under a specified variance model and a nonparametric alternative. Their theoretical development relies on a location-scale formulation, so their results do not directly address the broader constant-variance null with covariate-dependent higher-order error features. \citet{Zheng2009} proposes a quadratic kernel test for the conditional variance restriction and implements it using a wild bootstrap. Its local-alternative detection rate, however, is of order $n^{-1/2}h^{-q/4}$ \citep{Chown2018} and therefore depends on the bandwidth and covariate dimension, so the test does not attain root-$n$ sensitivity to the Pitman-type local alternatives considered here. \citet{DetteHetzler2009} develop a pseudo-residual empirical process with root-$n$ local sensitivity for testing a parametric variance function under fixed designs and extend their analysis to univariate random designs. Although their general asymptotic theory permits higher-order conditional moments to vary, their homoskedasticity calibration requires a constant conditional fourth moment: their Corollaries~7 and~10 (the latter under a random design) obtain a scaled Brownian bridge only when this moment is constant, restricting the applicability of their calibration under the broader null. These distinctions motivate the development of a cumulative variance test for random covariates with general fixed dimension, along with a calibration procedure that accommodates covariate-dependent conditional fourth moments.

To address these gaps, we construct Cram\'er--von Mises (CvM) and Kolmogorov--Smirnov (KS) test statistics from a third-order $U$-process based on cumulative conditional-moment restrictions. First, the proposed tests remain valid under covariate-dependent changes in conditional skewness, kurtosis, and other higher-order features of the error distribution, since validity requires only constancy of the conditional second moment rather than independence between the error and covariates. Second, the test statistics are based on a residual-marked empirical process rather than on a kernel quadratic-form test whose local detection rate depends on the bandwidth. Moreover, the tests can detect Pitman-type local alternatives approaching the null at the root-$n$ rate. Third, the multiplier bootstrap yields valid critical values when the conditional fourth moment varies with the covariates and avoids re-estimating the nonparametric regression in each bootstrap replication.

The Monte Carlo experiments evaluate the calibration of the tests under homoskedasticity with covariate-dependent higher-order error features and examine their power against different patterns of variance heterogeneity. The results illustrate how the form of the departure affects the relative sensitivity of the CvM and KS statistics.

The NCDS application illustrates the economic relevance of testing conditional variance. Much of the literature on the labor-market height premium has focused on differences in mean earnings. For example, \citet{CasePaxson2008} link the height premium to cognitive ability and occupational sorting. Our analysis extends this perspective by examining how earnings dispersion varies with height and how the patterns differ between weekly and hourly outcomes. Across height intervals, the dispersion of log weekly earnings declines substantially, whereas the dispersion of log hourly earnings changes much less. A decomposition of weekly earnings shows that differences in the dispersion of hours worked and the covariance between hourly earnings and hours help explain this contrast. These patterns indicate that weekly-earnings dispersion is not fully accounted for by hourly-pay dispersion alone.

Sample composition provides a second dimension. Since the proportion of men and women differs markedly across height intervals, pooled earnings dispersion reflects both within-sex variation and differences in sex-specific conditional means. The proposed tests reject homoskedasticity for pooled weekly earnings but do not reject it in either sex-specific subsample. These patterns are descriptive and do not identify a causal effect of height on earnings inequality. Together with the male hourly-earnings comparison discussed earlier, where the stronger invariant-error restriction is rejected but the constant-variance restriction is not, the results illustrate how the choice of earnings measure, the composition of the sample, and the hypothesis under examination matter for economic interpretation.

Section~\ref{sec:statistics} introduces the hypotheses and test statistics. Section~\ref{sec:asymptotic} develops the asymptotic theory and multiplier bootstrap approximation, while Section~\ref{sec:bootstrap} discusses practical implementation. Sections~\ref{sec:simulation} and \ref{sec:empirical} present the Monte Carlo experiments and empirical application. Section~\ref{sec:conclusion} concludes. The supplementary material contains proofs, additional simulation results, and further empirical analyses in Appendices A--C.

\section{Hypotheses and Test Statistics}\label{sec:statistics}

\subsection{The Hypotheses}\label{sec:statistics_hypothesis}

Let $(\Omega,\mathcal A,\mathbb P)$ be the probability space of the random vector $(Y,X^\top)^\top$, where $Y$ is scalar and $X$ is $\mathbb R^q$-valued for $q\geq 1$. Consider the nonparametric regression
\begin{equation}
Y=m(X)+\varepsilon, \label{np}
\end{equation}
where $m(X)=\mathbb{E}[Y|X]$ is the unknown conditional mean function, and $\varepsilon$ is an error term satisfying $\mathbb{E}[\varepsilon|X]=0$ almost surely ($a.s.$) by construction. 
We consider regressions with $0<\mathbb E[\varepsilon^2]<\infty$. The null hypothesis of constant conditional variance is
\begin{equation}
H_0: \mathbb{E}[\varepsilon^2|X]=\sigma^2\,\,\,a.s. \text{ for some constant }0<\sigma^2<\infty\label{null}
\end{equation}
Note that $\sigma^2$ is also the unconditional variance of $\varepsilon$ when the null $H_0$ holds. The alternative is that the conditional variance is not almost surely constant:
\begin{equation*}
H_1:\ \text{there does not exist }c\in(0,\infty)\text{ such that }\mathbb E[\varepsilon^2\mid X]=c\quad a.s.
\end{equation*}
We seek omnibus tests of this restriction in model \eqref{np}, allowing higher-order conditional features of $\varepsilon$ to depend on $X$. 

Under $H_0$ in \eqref{null}, $\varepsilon$ need not be independent of $X$. The conditional moments $\mathbb{E}[\varepsilon^3|X]$ and $\mathbb{E}[\varepsilon^4|X]$, when they exist, may vary with $X$ even when the conditional variance is constant. The location-scale model \eqref{lc} imposes additional restrictions on these moments under homoskedasticity. Procedures calibrated under such restrictions need not retain their nominal interpretation under the broader null \eqref{null}, while rejection of the stronger restriction alone does not by itself establish heteroskedasticity.

We use an empirical-process construction that does not require higher-order conditional moments to be constant. Suppose that $X$ has a Lebesgue density $f(\cdot)$. The null restriction
\begin{equation*}
\mathbb{E}\left[\varepsilon^2-\sigma^2|X\right]=0\,\,\,a.s.,
\end{equation*}
is equivalent to 
\begin{equation*}
f^2(X)\mathbb{E}\left[\varepsilon^2-\sigma^2|X\right]=0\,\,\,a.s.,
\end{equation*}
because $f(X)>0$ $a.s.$ The squared-density weight $f^2(X)$ avoids division by pointwise density estimates in the construction below. Equivalently, the null restriction can be expressed as 
\begin{equation}
T(x):=\mathbb{E}\left[f^2(X)\left(\varepsilon^2-\sigma^2\right)1(X\leq x)\right]=0\,\,\,\text{for all}\,\,\,x\in\mathbb R^q.\label{equivalence}
\end{equation}
Henceforth, $1(A)$ is the indicator function of the event $A$. For two vectors $u$ and $v$ of equal dimension, $u\leq v$ means that each coordinate of $u$ is less than or equal to the corresponding coordinate of $v$. The process $T(x)$ is zero for every $x$ under $H_0$. Under $H_1$, the same cumulative functional evaluated at the probability limit $\sigma_\ast^2$ defined in Section~\ref{sec:test_statistics} is nonzero on a set of $x$ with positive Lebesgue measure, providing the basis for our tests.

\subsection{Test Statistics}\label{sec:test_statistics}
Suppose that the sample consists of $n\geq 3$ independent and identically distributed (i.i.d.) observations $\left\{W_i=(Y_i,X_i^\top)^\top\right\}_{i=1}^n$ of $(Y,X^\top)^\top$. Define $\sigma_\ast^2 := \mathbb E[f^2(X)\varepsilon^2]/\mathbb E[f^2(X)]$ and notice that $\sigma^2=\sigma_\ast^2$ under $H_0$. A preliminary estimator of $\sigma^2$ is 
\begin{equation*}
\frac{\sum_{i=1}^n\hat f^2(X_i)\hat\varepsilon_i^2}{\sum_{i=1}^n\hat f^2(X_i)}=\frac{\sum_{i=1}^n\sum_{j\neq i}^n\sum_{k\neq i}^nK_{ij}K_{ik}(Y_i-Y_j)(Y_i-Y_k)}{\sum_{i=1}^n\sum_{j\neq i}^n\sum_{k\neq i}^nK_{ij}K_{ik}},
\end{equation*}
where $\hat{\varepsilon}_{i}=Y_i-\hat{m}(X_i)$ for $i=1,\ldots,n$ are the nonparametric residuals and
\[\hat{m}(X_i)=\frac{1}{\hat{f}(X_i)}\frac{1}{(n-1)a^q}\sum_{j\neq i}^{n}K_{ij}Y_j \text{ and } \hat{f}(X_i)=\frac{1}{(n-1)a^q}\sum_{j\neq i}^{n}K_{ij}\]
are the leave-one-out estimators of $m(X_i)$ and $f(X_i)$. Here and henceforth, $K_{ij}=K((X_i-X_j)/a)$, $K(u)=\prod_{j=1}^qk(u_j)$, $k$ is a univariate kernel, and $a=a(n)\in\mathbb{R}^+$ is a bandwidth parameter shrinking to zero at an appropriate rate as sample size $n\rightarrow\infty$. Removing the repeated-index terms, for which the two neighbor indices coincide, gives the following ratio of third-order $U$-statistic components, defined whenever its aggregate denominator is nonzero:
\begin{equation}\label{sigma_u_statistic}
\hat\sigma^2=\frac{\sum_{i=1}^n\sum_{j\neq i}^n\sum_{k\neq i,j}^nK_{ij}K_{ik}(Y_i-Y_j)(Y_i-Y_k)}{\sum_{i=1}^n\sum_{j\neq i}^n\sum_{k\neq i,j}^nK_{ij}K_{ik}}.
\end{equation}

Using $\hat\sigma^2$, a sample analogue of $T(x)$ in \eqref{equivalence} is
\begin{equation*}
\frac{1}{n}\sum_{i=1}^n\hat f^2(X_i)\left(\hat\varepsilon_i^2-\hat\sigma^2\right)1(X_i\leq x), 
\end{equation*}
which can be further decomposed as
\begin{align}
&\frac{1}{n(n-1)^2}\sum_{i=1}^n\sum_{j\neq i}^n\sum_{k\neq i}^n\frac{1}{a^{2q}}K_{ij}K_{ik}\left[(Y_i-Y_j)(Y_i-Y_k)-\hat\sigma^2\right]1(X_i\leq x)\nonumber\\
=&\frac{1}{n(n-1)^2}\sum_{i=1}^n\sum_{j\neq i}^n\sum_{k\neq i,j}^n\frac{1}{a^{2q}}K_{ij}K_{ik}\left[(Y_i-Y_j)(Y_i-Y_k)-\hat\sigma^2\right]1(X_i\leq x)\nonumber\\
&+\frac{1}{n(n-1)^2}\sum_{i=1}^n\sum_{j\neq i}^n\frac{1}{a^{2q}}K^2_{ij}\left[(Y_i-Y_j)^2-\hat\sigma^2\right]1(X_i\leq x).\label{aux_T}
\end{align}
In \eqref{aux_T}, the first term determines the asymptotic distribution, while the second is a bias term satisfying
\begin{equation*}
\sup_x\left\vert \frac{1}{n(n-1)^2}\sum_{i=1}^n\sum_{j\neq i}^n\frac{1}{a^{2q}}K^2_{ij}\left[(Y_i-Y_j)^2-\hat\sigma^2\right]1(X_i\leq x)\right\vert=O_p\left((na^q)^{-1}\right).
\end{equation*}
Following the construction of $\hat\sigma^2$ in \eqref{sigma_u_statistic}, we drop the bias term in \eqref{aux_T} and consider the following $U$-process-type construction:
\begin{equation}\label{Tn}
\hat T_n(x)=\frac{1}{n(n-1)(n-2)}\sum_{i=1}^n\sum_{j\neq i}^n\sum_{k\neq i,j}^n\frac{1}{a^{2q}}K_{ij}K_{ik}\left[(Y_i-Y_j)(Y_i-Y_k)-\hat\sigma^2\right]1(X_i\leq x).
\end{equation}
The normalization in \eqref{Tn} differs from that of the distinct-index term in \eqref{aux_T} by the factor $(n-1)/(n-2)=1+O(n^{-1})$, which is asymptotically negligible under the $n^{1/2}$ scaling used below.
This construction is related to the $U$-process methods studied by \citet{stute1994u} and \citet{DelgadoGM2001}. 

\begin{remark}\label{rk:sigma hat}
Under $H_0$, the estimator $\hat\sigma^2$ in \eqref{sigma_u_statistic} is first-order equivalent to the sample-weighted ratio
\begin{equation*}
\frac{\sum_{i=1}^n f^2(X_i)\varepsilon_i^2}{\sum_{i=1}^n f^2(X_i)},
\end{equation*}
in the sense that their difference is $o_p(n^{-1/2})$, as follows from Proposition~\ref{le1}.
Unlike conventional residual-based estimators in the literature, the construction avoids the observation-specific random-density denominators that arise from direct plug-in of nonparametric residuals. The aggregate denominator in \eqref{sigma_u_statistic} remains random, as discussed in Remark~\ref{rk:breve}.
\end{remark}

The statistic $\hat T_n(x)$ in \eqref{Tn} is a third-order $U$-process-type construction with the estimated nuisance parameter $\hat\sigma^2$. Its component kernels depend on $n$ through the bandwidth $a$. Section~A.1.1 of the supplementary material gives the corresponding decomposition. For second-order non-degenerate $U$-processes in nonparametric significance testing, see \citet{DelgadoGM2001}. Our proposed test statistics are continuous functionals of the random element $n^{1/2}\hat T_n(x)$. The Cram\'er--von Mises-type statistic is
\begin{equation*}
\widehat{CvM}_n=\int\left[n^{1/2}\hat T_n(x)\right]^2\,F_{Xn}(dx)=\sum_{i=1}^n\hat T_n(X_i)^2,
\end{equation*}
where, henceforth, $F_\zeta$ denotes the distribution function of the (possibly vector-valued) random variable $\zeta$ and $F_{\zeta n}$ denotes its corresponding empirical distribution function (EDF). The Kolmogorov--Smirnov-type test statistic is
\begin{equation*}
\widehat{KS}_n=\sup_x\left\vert n^{1/2}\hat T_n(x)\right\vert.
\end{equation*}
Both statistics aggregate the same cumulative variance discrepancy. CvM integrates its squared magnitude over the design distribution, whereas KS uses its largest absolute value. Calibrating either statistic requires accounting for the first-order effect of $\hat\sigma^2$ as well as estimation of the regression function.

\section{Asymptotic Theory}\label{sec:asymptotic}

Under the null, estimation of the common variance changes the centering of the first-order process. Its covariance depends on the conditional fourth moment, which may vary with the covariates. The resulting nonpivotal limit determines the calibration developed in Section~\ref{sec:bootstrap_theory}.

\subsection{Basic Assumptions}

Let $m_2(X)=\mathbb{E}[Y^2|X]$ and $\mu_4(X)=\mathbb E[\varepsilon^4\mid X]$. The fourth conditional moment $\mu_4(X)$ is allowed to vary with $X$. We use the kernel and smoothness--moment classes of \citet{Robinson1988}, with the following notation. \begin{samepage}
For $l\geq1$, let $\mathcal K_l$ be the class of even functions of uniformly bounded variation, $k:\mathbb R\to\mathbb R$, satisfying
\begin{equation*}
k(u)=O\left(\frac{1}{1+|u|^{l+1+\epsilon}}\right),\qquad \text{for some }\epsilon>0,
\end{equation*}
and
\begin{equation*}
\int_{\mathbb R}u^ik(u)\,du=\delta_{i0},\qquad i=0,\ldots,l-1,
\end{equation*}
where $\delta_{ij}$ is Kronecker's delta with $\delta_{ij}=1$ if $i=j$ and $\delta_{ij}=0$ otherwise.
\end{samepage}

Let $\alpha>0$ and $\beta>0$, and denote by $b=\lceil\beta\rceil$ the smallest integer not less than $\beta$. A measurable function $g:\mathbb R^q\to\mathbb R$ is said to belong to $\mathcal F_\beta^\alpha$ if it is $(b-1)$-times continuously differentiable and admits a local Taylor expansion of order $b-1$ with a remainder satisfying a H\"older-type bound. Specifically, there exist $\varrho>0$ and a measurable remainder function $D:\mathbb R^q\to[0,\infty)$ such that for every $u\in\mathbb R^q$ and all $v$ with $\|v-u\|<\varrho$,
\begin{equation*}
|g(v)-g(u)-Q(v,u)|\leq D(u)\|v-u\|^\beta.
\end{equation*}
Here $Q(v,u)=0$ when $b=1$. When $b>1$, $Q(v,u)$ is the nonconstant part of the $(b-1)$-th order Taylor polynomial of $g$ at $u$, given by
\begin{equation*}
Q(v,u)=\sum_{1\leq|\mathbf m|\leq b-1}\frac{\partial^{\mathbf m}g(u)}{\mathbf m!}(v-u)^{\mathbf m}.
\end{equation*}
Here $\mathbf m=(m_1,\ldots,m_q)$ is a multi-index and $|\mathbf m|=\sum_{j=1}^qm_j$. The integrability requirements in $\mathcal F_\beta^\alpha$ are understood with respect to the distribution of the design variable $X$. The function $g$, all its partial derivatives up to order $b-1$, and the remainder bound $D$ satisfy
\begin{equation*}
\mathbb E|g(X)|^\alpha<\infty,\qquad \mathbb E|\partial^{\mathbf m}g(X)|^\alpha<\infty\ \text{for }1\leq|\mathbf m|\leq b-1,\qquad \mathbb E|D(X)|^\alpha<\infty.
\end{equation*}
When $\alpha=\infty$, these requirements mean essential boundedness of the function, its relevant derivatives, and the remainder bound $D$ with respect to the law of $X$. Thus $\mathcal F_\beta^\alpha$ combines a smoothness index $\beta$ with an integrability index $\alpha$.

Following \citet{DelgadoGM2001}, we impose the following regularity conditions:

\begin{description}
\item Assumption 1. $f\in\mathcal F_\lambda^\infty$, for some $\lambda>q-1$.

\item Assumption 2. $m\in\mathcal F_\tau^4$, for some $\tau>q$.

\item Assumption 3. $k\in\mathcal K_{l+t-1}$, where $l-1<\lambda\leq l$ and $t-1<\tau\leq t$.

\item Assumption 4. $(na^{2q})^{-1}+na^{2\min(\lambda+1,\tau)}\rightarrow 0$ as $n\rightarrow\infty$.

\item Assumption 5. $\mathbb E|\varepsilon|^{4+\delta}<\infty$ for some $\delta>0$, where $\varepsilon=Y-m(X)$.

\item Assumption 6 (Nondegeneracy). Under $H_0$, $\mathbb E[(\varepsilon^2-\sigma^2)^2]>0$.

\end{description}

\begin{remark}[Nondegeneracy]\label{rk:nondegeneracy}
Assumption 6 excludes only the special case in which the squared error is almost surely constant. Under $H_0$,
\begin{equation*}
\mathbb E[(\varepsilon^2-\sigma^2)^2]
=\mathbb E\{\mu_4(X)-\sigma^4\},\qquad
\mu_4(X)-\sigma^4=\operatorname{Var}(\varepsilon^2\mid X)\geq0.
\end{equation*}
It therefore fails if and only if $\varepsilon^2=\sigma^2$ almost surely, as for a mean-zero error taking the two values $\pm\sigma$. The condition permits the fourth conditional moment to vary with $X$ and does not require a positive lower bound on the conditional variance of the squared error. It is not needed for identification, the uniform process approximations, or fixed-alternative consistency. Its role is to make the null Gaussian limit nontrivial and justify exact limiting critical values.
\end{remark}

\begin{remark}
Assumptions 1--3 are standard regularity conditions for nonparametric smoothing. Assumption 4 is an undersmoothing condition balancing the stochastic order of the higher-order terms in the Hoeffding decomposition and the smoothing bias. Equivalently, it requires $n^{-1/(2q)}\ll a\ll n^{-1/\{2\min(\lambda+1,\tau)\}}$, and this interval is nonempty because Assumptions 1 and 2 imply $\min(\lambda+1,\tau)>q$. Assumption 5 provides the moment condition required for the squared-error marked processes considered in this paper.
\end{remark}

Under $H_0$, the probability limit $\sigma_\ast^2$ of $\hat\sigma^2$ equals $\sigma^2$. The following proposition establishes the asymptotic properties of $\hat\sigma^2$ under the null hypothesis.

\begin{samepage}
\begin{proposition}\label{le1}
Suppose Assumptions 1--5 hold. Then, under $H_0$, we have that
\begin{equation*}
\hat\sigma^2-\sigma^2=\frac{1}{\mathbb{E}[f^2(X)]}\frac{1}{n}\sum_{i=1}^nf^2(X_i)\left(\varepsilon_i^2-\sigma^2\right)+o_p(n^{-1/2}). 
\end{equation*}
Consequently,
\begin{equation*}
\sqrt n\left(\hat\sigma^2-\sigma^2\right)\xrightarrow{d}N\left(0,\frac{\mathbb{E}\left[f^4(X)\left(\varepsilon^2-\sigma^2\right)^2\right]}{\left(\mathbb{E}[f^2(X)]\right)^2}\right). 
\end{equation*}
\end{proposition}

Recall that $\mu_4(X)=\mathbb E[\varepsilon^4\mid X]$. Proposition~\ref{le1} gives
\begin{equation}
\sqrt n\left(\hat\sigma^2-\sigma^2\right)\xrightarrow{d}N\left(0,\frac{\mathbb{E}\left[f^4(X)\left(\mu_4(X)-\sigma^4\right)\right]}{\left(\mathbb{E}[f^2(X)]\right)^2}\right). \label{sigma_normality_null}
\end{equation}
\end{samepage}

\begin{remark}\label{rk:breve}
The squared-density-weighted estimator $\hat\sigma^2$ in \eqref{sigma_u_statistic} avoids the observation-specific random density denominators in the conventional residual-based estimator $\breve\sigma^2=n^{-1}\sum_{i=1}^n\hat\varepsilon_i^2$. 
Its aggregate denominator is still random, and after division by $n(n-1)(n-2)a^{2q}$, it converges to $\mathbb E[f^2(X)]>0$. Under stronger assumptions, including bounded support of $X$, the residual-based estimator satisfies 
\begin{equation*}
\breve\sigma^2-\sigma^2=\frac{1}{n}\sum_{i=1}^n\left(\varepsilon_i^2-\sigma^2\right)+o_p(n^{-1/2}),
\end{equation*}
and thus
\begin{equation*}
\sqrt{n}\left(\breve\sigma^2-\sigma^2\right)\xrightarrow{d} N(0,\mathbb{E}[\varepsilon^4]-\sigma^4),
\end{equation*}
where $\mathbb{E}[\varepsilon^4]$ denotes the unconditional fourth moment of $\varepsilon$. To see the difference between $\hat\sigma^2$ and $\breve\sigma^2$, note that if $\mathbb{E}[\varepsilon^4|X]=\mathbb{E}[\varepsilon^4]$ holds under the null, that is, a constant fourth conditional moment (for example, when $\varepsilon\perp X$ is further assumed under the null), then
\begin{equation*}
\sqrt n\left(\hat\sigma^2-\sigma^2\right)\xrightarrow{d}N\left(0,\frac{\mathbb{E}\left[f^4(X)\right]}{\left(\mathbb{E}[f^2(X)]\right)^2}\left(\mathbb{E}[\varepsilon^4]-\sigma^4\right)\right).
\end{equation*}
By Jensen's inequality,
\begin{equation*}
\mathbb{E}\left[f^4(X)\right]\geq\left(\mathbb{E}\left[f^2(X)\right]\right)^2,
\end{equation*}
implying that $\hat\sigma^2$ is not as efficient as $\breve\sigma^2$, which has asymptotic variance $\mathbb{E}[\varepsilon^4]-\sigma^4$. Nevertheless, $\hat\sigma^2$ requires less restrictive assumptions and is suitable for our testing purpose. In addition, using an estimator with smaller asymptotic variance, such as $\breve\sigma^2$ under a constant fourth conditional moment, need not yield a more powerful test.

A general class of density-weighted estimators is discussed in Section~A.6.1 of the supplementary material.
\end{remark}

\subsection{Asymptotic Null Distributions}

\begin{samepage}
To derive the null limit of the test process, define
\begin{equation*}
r(x)=\frac{\mathbb E[f^2(X)1(X\leq x)]}{\mathbb E[f^2(X)]}.
\end{equation*}
Write
\begin{equation*}
T_n^0(x)=\frac{1}{n}\sum_{i=1}^n\xi_x(W_i),
\end{equation*}
where
\begin{equation*}
\xi_x(W_i)=f^2(X_i)\left(\varepsilon_i^2-\sigma^2\right)\left[1(X_i\leq x)-r(x)\right].
\end{equation*}
\end{samepage}
Under $H_0$, $\mathbb E[\xi_x(W)]=0$ and
\begin{equation*}
\mathbb E[\xi_x(W)^2]
=\mathbb E\left[f^4(X)\{\mu_4(X)-\sigma^4\}
\{1(X\leq x)-r(x)\}^2\right].
\end{equation*}
Under Assumption 6, this variance is positive whenever $0<r(x)<1$, and such indices exist. Thus the Gaussian limit below is nontrivial. The weak-convergence results require only Assumptions 1--5.

The following theorem establishes uniform asymptotic equivalence between $\hat T_n$ and $T_n^0$.

\begin{theorem}\label{th1}
Suppose Assumptions 1--5 hold. Then, under $H_0$, we have that
\begin{align*}
\sup_x\left\vert \hat T_n(x)-T_n^0(x)\right\vert=o_p(n^{-1/2}).
\end{align*}
\end{theorem}

By Theorem \ref{th1}, the null limit of $n^{1/2}\hat T_n$ is determined by $n^{1/2}T_n^0$. We therefore view $T_n^0$ as indexed by the function class $\mathcal H:=\{\xi_x:x\in\mathbb{R}^q\}$.

\begin{samepage}
\begin{corollary}\label{IT}
Suppose Assumptions 1--5 hold. Then, under $H_0$, we have that
\begin{equation*}
n^{1/2}\hat T_n\Rightarrow B_{\mathbb P}\,\,\,\text{in}\,\,\,\ell^\infty(\mathcal H),
\end{equation*}
where $B_{\mathbb P}$ is a centered tight Gaussian process indexed by $\mathcal H$ with covariance $\mathbb E[B_{\mathbb P}gB_{\mathbb P}h]=\mathbb P gh$ for $g,h\in\mathcal H$. Furthermore,
\begin{equation*}
\widehat{CvM}_n\xrightarrow{d}\int\left\vert B_{\mathbb P}\xi_x\right\vert^2\,F_{X}(dx),
\end{equation*}
\begin{equation*}
\widehat{KS}_n\xrightarrow{d}\sup_x\left\vert B_{\mathbb P}\xi_x\right\vert.
\end{equation*}

\end{corollary}
\end{samepage}

\begin{remark}[Covariance and Exact Quantiles]\label{rk:covkernel}
The covariance kernel of the limiting Gaussian process $B_{\mathbb P}$ is
\begin{equation*}
\mathbb E\left[B_{\mathbb P}\xi_xB_{\mathbb P}\xi_z\right]
=\mathbb E\Big[f^4(X)\big(\mu_4(X)-\sigma^4\big)\big\{1(X\le x)-r(x)\big\}\big\{1(X\le z)-r(z)\big\}\Big].
\end{equation*}
The kernel depends on $f$, $r$, the fourth conditional moment $\mu_4$, and the null variance $\sigma^2$, so the null distributions are nonpivotal. Under Assumption 6, the KS limit has a continuous, strictly increasing distribution on its positive support: Theorem~3 of \citet{Yang2025Continuity} covers the case of zero minimum variance, and Theorem~2 of \citet{Yang2025Strict} gives strict monotonicity under compactness and a full-support measure. No positive lower bound on the pointwise variances is required. The CvM limit is a nondegenerate weighted sum of independent chi-squared variables and also has a continuous, strictly increasing distribution on $(0,\infty)$. Both limits therefore have unique exact $(1-\alpha)$-quantiles for $\alpha\in(0,1)$. Section~A.5 of the supplementary material verifies these assertions.
\end{remark}

\subsection{Asymptotic Power under Alternatives}\label{sec:local_power}

Under $H_1$, with $T(x)$ defined using the probability limit $\sigma_\ast^2$ of $\hat\sigma^2$, the function
\begin{equation*}
T(x)=\mathbb E\left[f^2(X)\left(\varepsilon^2-\sigma_\ast^2\right)1(X\leq x)\right]
\end{equation*}
is nonzero on a set of $x$ with positive Lebesgue measure. The following theorem establishes the uniform convergence of $\hat T_n(x)$ to $T(x)$, and its corollary establishes the consistency of the CvM and KS tests.

\begin{theorem}\label{global_theorem}
Suppose Assumptions 1--5 hold. Then, under $H_{1}$, we have that
\begin{align*}
\sup_x\left\vert \hat T_n(x)-T(x)\right\vert=o_p(1).
\end{align*}
\end{theorem}
\begin{samepage}
\begin{corollary}
Suppose Assumptions 1--5 hold. Then, under $H_{1}$, we have that
\begin{equation*}
\frac{\widehat{CvM}_n}{n}\xrightarrow{p}\int\left\vert T(x)\right\vert^2\,F_{X}(dx)>0,
\end{equation*}
\begin{equation*}
\frac{\widehat{KS}_n}{\sqrt n}\xrightarrow{p}\sup_x\left\vert T(x)\right\vert>0.
\end{equation*}
\end{corollary}

\end{samepage}

Consider the Pitman sequence with $\mu_4(X)=\mathbb E[\varepsilon^4|X]$ kept fixed:
\begin{equation}
H_{1n}: \mathbb{E}[\varepsilon^2|X]=\sigma^2+\frac{\pi(X)}{\sqrt n}\,\,\,a.s., \label{pitman}
\end{equation}
where the local deviation $\pi(\cdot)$ is a nonzero function defined on $\mathbb R^q$ such that $\mathbb E[\pi(X)]=0$, $0<\mathbb E[\pi^2(X)]<\infty$, and the conditional variance $\mathbb E[\varepsilon^2\mid X]$ remains nonnegative. The regularity conditions in Assumptions 1--5 are understood to hold uniformly along the local sequence. To keep the notation light, the row index $n$ is suppressed in $\varepsilon_i$ and related expressions.

Under $H_{1n}$ in \eqref{pitman}, $\hat\sigma^2$ admits the following asymptotic expansion:
\begin{equation*}
\hat\sigma^2-\sigma^2=\frac{1}{\sqrt n}\frac{\mathbb{E}[f^2(X)\pi(X)]}{\mathbb{E}[f^2(X)]}+\frac{1}{\mathbb{E}[f^2(X)]}\frac{1}{n}\sum_{i=1}^nf^2(X_i)\left(\varepsilon_i^2-\sigma_\ast^2\right)+o_p(n^{-1/2}),
\end{equation*}
and consequently,
\begin{equation*}
\sqrt n\left(\hat\sigma^2-\sigma^2\right)\xrightarrow{d}N\left(\frac{\mathbb{E}[f^2(X)\pi(X)]}{\mathbb{E}[f^2(X)]},\frac{\mathbb{E}\left[f^4(X)\left(\mu_4(X)-\sigma^4\right)\right]}{\left(\mathbb{E}[f^2(X)]\right)^2}\right).
\end{equation*}
Thus $\hat\sigma^2$ remains consistent under $H_{1n}$, while $\sqrt n(\hat\sigma^2-\sigma^2)$ may have a nonzero limiting mean, given by $\mathbb{E}[f^2(X)\pi(X)]/\mathbb{E}[f^2(X)]$. Its asymptotic variance is the expression in \eqref{sigma_normality_null}.

Let
\begin{align*}
    \Pi(x) = \mathbb E \left\{ f^2(X) \left[\pi(X)- \frac{\mathbb E \left[ f^2(X)\pi(X) \right]}{\mathbb E \left[ f^2(X) \right]}\right]1\left(X\le x\right)\right\}
\end{align*}
be a deterministic function that is nonzero on a set of $x$ with positive Lebesgue measure under $H_{1n}$. Estimating the common variance absorbs the density-weighted average of the local deviation, leaving the centered deviation inside this cumulative expectation. Consequently, local power depends on its accumulation across lower orthants, not simply on the pointwise magnitude of the change in variance. Departures with different patterns of sign changes can therefore induce different drift functions even when their pointwise magnitudes are comparable. The following theorem establishes that $n^{1/2}\hat T_n$ converges weakly to a centered Gaussian process plus the deterministic drift function $\Pi$ under the local alternatives $H_{1n}$.

\begin{theorem}\label{local_theorem}
Suppose Assumptions 1--5 hold. Then, under $H_{1n}$, we have that
\begin{equation*}
n^{1/2}\hat T_n\Rightarrow B_{\mathbb P}+\Pi\,\,\,\text{in}\,\,\,\ell^\infty(\mathcal H),
\end{equation*}
where $B_{\mathbb P}$ is the same centered Gaussian process defined in Corollary \ref{IT}. 
\end{theorem}

Theorem \ref{local_theorem} yields the following limits for $\widehat{CvM}_n$ and $\widehat{KS}_n$, which show their sensitivity to alternatives that approach $H_0$ at the parametric rate.
\begin{samepage}
\begin{corollary}\label{coro_local}
Suppose Assumptions 1--5 hold. Then, under $H_{1n}$, we have that
\begin{equation*}
\widehat{CvM}_n\xrightarrow{d}\int\left\vert B_{\mathbb P}\xi_x+\Pi(x)\right\vert^2\,F_{X}(dx),
\end{equation*}
\begin{equation*}
\widehat{KS}_n\xrightarrow{d}\sup_x\left\vert B_{\mathbb P}\xi_x+\Pi(x)\right\vert.
\end{equation*}
\end{corollary}
\end{samepage}

\Needspace{12\baselineskip}
\subsection{Multiplier Bootstrap Approximation}\label{sec:bootstrap_theory}
The limiting null distributions are nonpivotal and, under Assumption 6, nontrivial, so we use a multiplier bootstrap to approximate the asymptotic behavior of the test statistics under $H_0$. Define 
\begin{equation*}
\hat r(x)=\frac{\sum_{i=1}^n\sum_{j\neq i}^n\sum_{k\neq i,j}^nK_{ij}K_{ik}1(X_i\leq x)}{\sum_{i=1}^n\sum_{j\neq i}^n\sum_{k\neq i,j}^nK_{ij}K_{ik}},
\end{equation*}
which is a uniformly consistent $U$-statistic-type estimator for $r(x)$. Then a possible multiplier bootstrap version of $\hat T_n(x)$ is given by
\begin{align*}
\frac{1}{n}\sum_{i=1}^nV_i\hat f^2(X_i)\left(\hat\varepsilon_i^2-\hat\sigma^2\right)\left[1(X_i\leq x)-\hat r(x)\right],
\end{align*}
where $\left\{V_i\right\}_{i=1}^n$ is a sequence of bounded i.i.d. random variables with mean zero and variance one and independent of $\left\{W_i=(Y_i,X_i^\top)^\top\right\}_{i=1}^n$. One choice is the two-point distribution of \citet{Mammen1993}, with $\mathbb{P}\left(V_i=(1+\sqrt{5})/2\right) =(\sqrt{5}-1)/(2\sqrt{5})$ and $\mathbb{P}\left(V_i=(1-\sqrt{5})/2\right) =(\sqrt{5}+1)/(2\sqrt{5})$. Alternatively, Rademacher multipliers satisfy $\mathbb{P}\left(V_i=1\right)=\mathbb{P}\left(V_i=-1\right) =1/2$. See, e.g., \cite{Liu1988} and \cite{de1996bierens}. 

Section~A.6.2 of the supplementary material gives the repeated-index decomposition of this expression and bounds the resulting bias term. To match the construction in \eqref{Tn}, we use the bias-corrected multiplier process
\begin{align*}
\hat T_n^\ast(x)
={}&\frac{1}{n(n-1)(n-2)}\sum_{i=1}^n\sum_{j\neq i}^n\sum_{k\neq i,j}^n
V_i\frac{1}{a^{2q}}K_{ij}K_{ik}\\
&\quad\times\left[(Y_i-Y_j)(Y_i-Y_k)-\hat\sigma^2\right]
\left[1(X_i\leq x)-\hat r(x)\right].
\end{align*}

\begin{samepage}
The corresponding bootstrap statistics for $\widehat{CvM}_n$ and $\widehat{KS}_n$ are
\begin{equation*}
\widehat{CvM}_n^\ast=\int\left[n^{1/2}\hat T_n^\ast(x)\right]^2\,F_{Xn}(dx)=\sum_{i=1}^n\hat T_n^\ast(X_i)^2,
\end{equation*}
and 
\begin{equation*}
\widehat{KS}_n^\ast=\sup_x\left\vert n^{1/2}\hat T_n^\ast(x)\right\vert.
\end{equation*}
\end{samepage}

Write
\begin{equation*}
T_n^{0\ast}(x)=\frac{1}{n}\sum_{i=1}^nV_i\xi_x(W_i).
\end{equation*}
The next theorem establishes that $n^{1/2}\hat T_n^\ast$ is asymptotically equivalent to $n^{1/2}T_n^{0\ast}$.

\begin{theorem}\label{th2}
Suppose Assumptions 1--5 hold. Then, under $H_0$ or $H_{1n}$, we have that
\begin{equation*}
\sup_x\left\vert\hat T_n^\ast(x)-T_n^{0\ast}(x)\right\vert=o_p(n^{-1/2}).
\end{equation*}
\end{theorem}

The following corollary gives the limiting bootstrap distributions conditional on the original sample $\left\{W_i=(Y_i,X_i^\top)^\top\right\}_{i=1}^n$. Their quantiles provide critical values. 

\begin{corollary}\label{coro2}
Suppose Assumptions 1--5 hold. Then, under $H_0$ or $H_{1n}$, we have that
\begin{equation*}
\widehat{CvM}_n^\ast\underset{\ast}{\overset{d}{\rightarrow}}\int\left\vert B_{\mathbb P}\xi_x\right\vert^2\,F_{X}(dx)\,\,\,\text{in probability},
\end{equation*}
and
\begin{equation*}
\widehat{KS}_n^\ast\underset{\ast}{\overset{d}{\rightarrow }}\sup_x\left\vert B_{\mathbb P}\xi_x\right\vert\,\,\,\text{in probability}.
\end{equation*}

Moreover, under $H_1$,
\begin{equation*}
n^{1/2}\hat T_n^\ast\underset{\ast}{\overset{d}{\rightarrow}}B_{\mathbb P}^{(1)}\,\,\,\text{in}\,\,\,\ell^\infty(\mathcal H^{(1)})\,\,\,\text{in probability},
\end{equation*}
where $\mathcal H^{(1)}=\{\xi_x^{(1)}:x\in\mathbb R^q\}$,
\begin{equation*}
\xi_x^{(1)}(W)=f^2(X)\left(\varepsilon^2-\sigma_\ast^2\right)\left[1(X\leq x)-r(x)\right],
\end{equation*}
and $B_{\mathbb P}^{(1)}$ is a centered Gaussian process with covariance function
\begin{equation*}
\mathbb E\left[B_{\mathbb P}^{(1)}\xi_x^{(1)}B_{\mathbb P}^{(1)}\xi_z^{(1)}\right]
=
\mathbb E\left[\xi_x^{(1)}(W)\xi_z^{(1)}(W)\right].
\end{equation*}
\begin{samepage}
Consequently,
\begin{equation*}
\widehat{CvM}_n^\ast\underset{\ast}{\overset{d}{\rightarrow}}
\int\left\vert B_{\mathbb P}^{(1)}\xi_x^{(1)}\right\vert^2\,F_X(dx)\,\,\,\text{in probability},
\end{equation*}
and
\begin{equation*}
\widehat{KS}_n^\ast\underset{\ast}{\overset{d}{\rightarrow}}
\sup_x\left\vert B_{\mathbb P}^{(1)}\xi_x^{(1)}\right\vert\,\,\,\text{in probability}.
\end{equation*}
\end{samepage}
Here $\underset{\ast }{\overset{d}{\rightarrow }}$ denotes the convergence in distribution under the bootstrap law, i.e., conditional on the original sample $\left\{W_i=(Y_{i},X_{i}^{\top})^{\top}\right\}_{i=1}^{n}$.
\end{corollary}

\begin{remark}[Bootstrap Critical Values]\label{rk:size}
Suppose Assumptions 1--5 hold and, under $H_0$, impose Assumption 6. For either statistic, let $c_{1-\alpha}$ be the unique exact $(1-\alpha)$-quantile of its null limit, as established in Remark~\ref{rk:covkernel}, and let $c^\ast_{n,1-\alpha}$ be the corresponding conditional bootstrap quantile. Corollary~\ref{coro2}, together with continuity and strict increase of the limiting distribution, gives
\begin{equation*}
c^\ast_{n,1-\alpha}\xrightarrow{p}c_{1-\alpha}.
\end{equation*}
Rejecting when the observed statistic is at least $c^\ast_{n,1-\alpha}$ therefore has asymptotic size $\alpha$ under $H_0$.

Under $H_{1n}$, impose Assumption 6 on the limiting null law. The bootstrap critical value converges to the same oracle null critical value. If the corresponding local limiting statistic in Corollary~\ref{coro_local} has no atom at that critical value, the bootstrap test has the same first-order local power as the oracle test. Under fixed alternatives, Assumptions 1--5 suffice because the bootstrap statistic is $O_p(1)$ by Corollary~\ref{coro2}, whereas the original statistic diverges at the rate given in Section~\ref{sec:local_power}, so the bootstrap test is consistent. The computational steps are described in Section~\ref{sec:bootstrap}.
\end{remark}

\Needspace{10\baselineskip}
\section{Practical Implementation}\label{sec:bootstrap}

For the main Monte Carlo experiments and the NCDS application, which involve a scalar covariate, we use the Epanechnikov kernel
\begin{equation*}
k(u)=\frac{3}{4}(1-u^2)1(|u|\leq1)
\end{equation*}
and set
\begin{equation*}
a=C s_X n^{-1/3},\qquad
s_X=\min\left\{\operatorname{sd}(X),\frac{\operatorname{IQR}(X)}{1.349}\right\}.
\end{equation*}
The scale $s_X$ is calculated from the observed covariates. Following pilot comparisons of empirical size, we use $C=4$ for the main tests. This choice was assessed using independent simulation replications, and $C=3.5$ and $C=4.5$ are used to examine bandwidth sensitivity. The two-dimensional feasibility experiment in Section~B.5 of the supplementary material instead uses a fourth-order product kernel with coordinatewise bandwidths $a_j=C s_j n^{-1/5}$ (where $s_j$ is the coordinatewise analog of $s_X$) and $C=1.75$. These bandwidth choices are consistent with Assumption~4 under the data-generating settings considered.

Given the bandwidth, calculate the kernel weights $K_{ij}=K((X_i-X_j)/a)$ for $i\ne j$. These weights yield the common-variance estimator $\hat\sigma^2$ in \eqref{sigma_u_statistic} and the centering function $\hat r(x)$ in Section~\ref{sec:bootstrap_theory}. We then compute the process $\hat T_n(x)$ in \eqref{Tn} and evaluate the observed statistics $\widehat{CvM}_n$ and $\widehat{KS}_n$. The CvM integral is evaluated against the empirical distribution of the observed covariates.

To obtain critical values, we use independent Rademacher multipliers with $\mathbb P(V_i=1)=\mathbb P(V_i=-1)=1/2$. For each replication $b=1,\ldots,B$, generate a new vector $(V_1^{(b)},\ldots,V_n^{(b)})$ and calculate the multiplier process $\hat T_n^{\ast(b)}(x)$ and its CvM and KS statistics as defined in Section~\ref{sec:bootstrap_theory}. The same multiplier vector is used for all thresholds $x$ within a replication, while the kernel weights, $\hat\sigma^2$, and $\hat r(x)$ are calculated once from the original sample. Thus each replication requires only a new multiplier draw, without re-estimating the regression function. We take $B=1{,}000$ in the Monte Carlo experiments and $B=5{,}000$ in the NCDS application. At significance level $\alpha$, the empirical $(1-\alpha)$-quantile of the $B$ bootstrap CvM or KS statistics provides the corresponding critical value. We reject $H_0$ when the observed statistic is at least this critical value.

\begin{remark}[Computational Reduction]
The distinct-index sums used in \eqref{sigma_u_statistic} and \eqref{Tn} can be calculated from pairwise kernel quantities. For each fixed $i$, we have
\begin{align*}
&\sum_{j\ne i}\sum_{k\ne i,j}K_{ij}K_{ik}(Y_i-Y_j)(Y_i-Y_k)\\
&\qquad=\left\{\sum_{j\ne i}K_{ij}(Y_i-Y_j)\right\}^{2}
-\sum_{j\ne i}K_{ij}^{2}(Y_i-Y_j)^{2},\\
&\sum_{j\ne i}\sum_{k\ne i,j}K_{ij}K_{ik}
=\left\{\sum_{j\ne i}K_{ij}\right\}^{2}-\sum_{j\ne i}K_{ij}^{2}.
\end{align*}
These exact identities remove the coincident-index terms $j=k$ without enumerating triples. The resulting pairwise quantities also enter the multiplier process, with the multiplier and threshold-dependent terms applied at the outer index $i$.
\end{remark}

\Needspace{10\baselineskip}
\section{Monte Carlo Experiments}\label{sec:simulation}

The simulations evaluate empirical size under constant conditional variance with covariate-dependent higher-order error features and compare power across different shapes and magnitudes of variance departures. All tests are evaluated at the $5\%$ level using $M=1{,}000$ Monte Carlo replications and $B=1{,}000$ bootstrap draws.

\subsection{Design}\label{sec:simulation_design}

For the main experiments, $n\in\{50,100,200\}$ and
\begin{equation*}
X\sim U(0,1),\qquad m(x)=1+\sin(2\pi x),\qquad
Y=m(X)+\sqrt{v_\rho(X)}e,
\end{equation*}
where $\mathbb E(e\mid X)=0$, $\mathbb E(e^2\mid X)=1$, and $\rho\in\{0,0.25,0.50,0.75\}$. We consider four designs.
\begin{enumerate}
\item \textbf{Gaussian, linear variance:} $e\sim N(0,1)$ independently of $X$ and $v_\rho(x)=1+\rho(2x-1)$.
\item \textbf{Gaussian, cosine variance:} the error distribution is as in Case 1 and $v_\rho(x)=1+\rho\cos(2\pi x)$.
\item \textbf{Conditional-$t$, linear variance:} conditional on $X=x$, let $Z_x\sim t_{\nu(x)}$ with $\nu(x)=10+6x$, and define $e=\{(\nu(X)-2)/\nu(X)\}^{1/2}Z_X$. The conditional variance of $e$ is one while its conditional kurtosis changes with $X$.
\Needspace{9\baselineskip}
\item \textbf{Skew-normal, linear variance:} conditional on $X=x$, let $Z_x\sim \mathrm{SN}(0,1,\alpha(x))$ with $\alpha(x)=1+4x$. Writing $\Delta(x)=\alpha(x)/\sqrt{1+\alpha(x)^2}$, define
\begin{equation*}
e=\frac{Z_X-\Delta(X)\sqrt{2/\pi}}
{\sqrt{1-2\Delta(X)^2/\pi}}.
\end{equation*}
The resulting error has conditional mean zero and variance one, while conditional skewness and kurtosis vary with $X$.
\end{enumerate}
Cases 3 and 4 use the linear variance function of Case 1. At $\rho=0$, their conditional variance is constant while their higher-order conditional features vary with $X$. Cases 1 and 2 share the same Gaussian null, which is evaluated only once at $\rho=0$. The uniform design is used as a standard bounded-support finite-sample benchmark.

The proposed tests are implemented as described in Section~\ref{sec:bootstrap}, with $C=4$ for the main results and $C=3.5$ and $C=4.5$ for bandwidth sensitivity checks.

We compare the proposed tests with \citet{Zheng2009} and with a smooth-residual-bootstrap implementation of the weighted-empirical-process procedure of \citet{Chown2018}, denoted CM-adapted. The latter uses bootstrap calibration rather than the original distribution-free calibration in light of \citet{Chown2019}. Cases 3 and 4 fall outside the standardized-error-independence null maintained by CM-adapted, and hence outside the location-scale model required by \citet{Chown2018}, so its rejection frequencies in those designs are reported for comparison rather than interpreted as size estimates.

\subsection{Size}\label{sec:simulation_size}

Table~\ref{tab:sim_size} reports rejection frequencies under constant conditional variance with $C=4$. Across the nine sample-size/error-distribution combinations, CvM ranges from $4.0\%$ to $5.9\%$ and KS from $4.0\%$ to $6.5\%$. The rejection frequencies of the proposed tests remain close to the nominal level in these designs when conditional skewness or kurtosis varies with $X$. The largest deviation among the proposed tests is the $6.5\%$ KS rejection rate under skew-normal errors at $n=100$, which falls to $5.7\%$ at $n=200$.

\begin{table}[!htbp]
\centering\small
\caption{Rejection frequencies under constant conditional variance at the 5\% level}\label{tab:sim_size}
\begin{tabular}{llrrrr}
\toprule
$n$ & Error distribution & CvM & KS & CM-adapted & Zheng \\
\midrule
50  & Gaussian        & 4.1 & 4.5 & 5.5 & 4.2 \\
50  & Conditional $t$ & 4.3 & 4.1 & 5.7$^{\dagger\ddagger}$ & 3.8 \\
50  & Skew-normal     & 4.4 & 5.4 & 7.6$^{\ddagger}$ & 4.8 \\
100 & Gaussian        & 4.0 & 4.9 & 6.3 & 3.7 \\
100 & Conditional $t$ & 4.6 & 4.9 & 6.6$^{\ddagger}$ & 3.5 \\
100 & Skew-normal     & 5.9 & 6.5 & 9.0$^{\ddagger}$ & 5.3 \\
200 & Gaussian        & 5.5 & 5.0 & 4.6 & 4.8 \\
200 & Conditional $t$ & 4.1 & 4.0 & 5.6$^{\ddagger}$ & 3.8 \\
200 & Skew-normal     & 5.8 & 5.7 & 8.2$^{\ddagger}$ & 6.5 \\
\bottomrule
\end{tabular}
\begin{minipage}{0.94\linewidth}\footnotesize
\emph{Notes:} Entries are percentages based on $M=B=1{,}000$. The proposed tests use $C=4$. Cases 1 and 2 share the Gaussian null. $\dagger$ The CM-adapted entry uses 998 valid replications. $\ddagger$ Conditional-$t$ and skew-normal designs lie outside the standardized-error-independence null of CM-adapted; the $\ddagger$-marked entries are therefore not size estimates for that procedure.
\end{minipage}
\end{table}

At $n=100$, the largest within-design spread across $C\in\{3.5,4,4.5\}$ is $0.5$ percentage points for CvM and $0.4$ for KS. Complete bandwidth sensitivity and Monte Carlo intervals are reported in Section~B.3 of the supplementary material.

\subsection{Power}\label{sec:simulation_power}

Tables~\ref{tab:sim_power100} and \ref{tab:sim_power200} report power at $n=100$ and $n=200$. Under the linear variance alternatives, the rejection frequencies of the proposed tests increase with both $\rho$ and $n$. At $n=200$ and $\rho=0.75$, CvM rejection frequencies are $95.2\%$, $91.8\%$, and $91.1\%$ in Cases 1, 3, and 4, respectively. The latter two cases show that the proposed tests retain power against variance changes even when higher-order conditional features also vary with the covariate.

The ranking changes under the cosine variance function in Case 2. At $n=200$ and $\rho=0.75$, the rejection frequencies for CvM, KS, CM-adapted, and Zheng are $68.6\%$, $82.3\%$, $72.5\%$, and $98.9\%$, respectively. This difference is consistent with the cumulative form of the population target. Under the uniform design,
\begin{equation*}
T_\rho(x)=\int_0^x\{v_\rho(t)-1\}\,dt.
\end{equation*}
For the linear and cosine alternatives,
\begin{equation*}
T_{\rho,\mathrm{lin}}(x)=\rho(x^2-x),\qquad
T_{\rho,\mathrm{cos}}(x)=\frac{\rho}{2\pi}\sin(2\pi x),
\end{equation*}
so
\begin{equation*}
\int_0^1T_{\rho,\mathrm{lin}}(x)^2dx=\frac{\rho^2}{30},\qquad
\int_0^1T_{\rho,\mathrm{cos}}(x)^2dx=\frac{\rho^2}{8\pi^2}.
\end{equation*}
The positive and negative portions of the cosine departure partially cancel in the cumulative target, which helps explain the lower CvM power in Case 2. This is the same cumulative feature that governs the local drift in Section~\ref{sec:local_power}, although the calculations here concern the fixed simulation alternatives. Consistent with this calculation, CvM power at $\rho=0.75$ rises from $28.4\%$ at $n=100$ to $68.6\%$ at $n=200$.

\begin{table}[!htbp]
\centering\small
\caption{Power at $n=100$}\label{tab:sim_power100}
\begin{tabular}{lrrrrr}
\toprule
Case & $\rho$ & CvM & KS & CM-adapted & Zheng \\
\midrule
1 & 0.25 & 10.6 & 9.2  & 5.2  & 8.0 \\
1 & 0.50 & 33.7 & 26.6 & 8.6  & 19.4 \\
1 & 0.75 & 65.7 & 52.7 & 22.3 & 48.1 \\
2 & 0.25 & 5.7  & 8.0  & 7.0  & 9.4 \\
2 & 0.50 & 12.8 & 23.4 & 14.5 & 31.7 \\
2 & 0.75 & 28.4 & 47.7 & 34.6 & 72.0 \\
3 & 0.25 & 12.1 & 9.5  & 7.8  & 7.0 \\
3 & 0.50 & 30.8 & 25.5 & 12.0 & 18.5 \\
3 & 0.75 & 59.1 & 50.8 & 22.9 & 43.2 \\
4 & 0.25 & 12.5 & 13.0 & 12.9 & 9.6 \\
4 & 0.50 & 34.1 & 30.0 & 23.9 & 21.7 \\
4 & 0.75 & 62.1 & 52.7 & 41.6 & 48.3 \\
\bottomrule
\end{tabular}
\begin{minipage}{0.94\linewidth}\footnotesize
\emph{Notes:} Entries are rejection frequencies in percent; $M=B=1{,}000$. The proposed tests use $C=4$. Cases 3 and 4 allow the standardized error distribution to vary with $X$; the CM-adapted entries in these cases are reported for comparison because its null is not maintained.
\end{minipage}
\end{table}

\begin{table}[!htbp]
\centering\small
\caption{Power at $n=200$}\label{tab:sim_power200}
\begin{tabular}{lrrrrr}
\toprule
Case & $\rho$ & CvM & KS & CM-adapted & Zheng \\
\midrule
1 & 0.25 & 20.9 & 17.5 & 7.1  & 14.2 \\
1 & 0.50 & 63.6 & 53.2 & 20.4 & 45.4 \\
1 & 0.75 & 95.2 & 88.0 & 52.6 & 87.2 \\
2 & 0.25 & 9.2  & 15.3 & 8.7  & 18.8 \\
2 & 0.50 & 27.4 & 45.0 & 28.8 & 70.4 \\
2 & 0.75 & 68.6 & 82.3 & 72.5 & 98.9 \\
3 & 0.25 & 18.5 & 15.5 & 7.9  & 10.5 \\
3 & 0.50 & 59.7 & 48.4 & 21.5 & 39.4 \\
3 & 0.75 & 91.8 & 83.4 & 52.0 & 78.9 \\
4 & 0.25 & 18.4 & 17.0 & 16.8 & 12.1 \\
4 & 0.50 & 58.6 & 51.4 & 39.2 & 40.4 \\
4 & 0.75 & 91.1 & 84.6 & 69.8 & 80.6 \\
\bottomrule
\end{tabular}
\begin{minipage}{0.94\linewidth}\footnotesize
\emph{Notes:} Entries are rejection frequencies in percent; $M=B=1{,}000$. The proposed tests use $C=4$. Cases 3 and 4 allow the standardized error distribution to vary with $X$; the CM-adapted entries in these cases are reported for comparison because its null is not maintained.
\end{minipage}
\end{table}

Taken together, the experiments show that the proposed tests maintain rejection frequencies close to the nominal level in the null designs with varying higher-order conditional features, while their power relative to the comparison procedures depends on the shape of the variance departure. Sections~B.2--B.6 of the supplementary material report power at $n=50$, complete bandwidth sensitivity, a strong-skewness Gamma stress test, a two-dimensional feasibility experiment, and an additional null design with a discontinuous fourth conditional moment.

\section{Empirical Application}\label{sec:empirical}

Height has been extensively studied as a correlate of mean earnings, but changes in average earnings do not by themselves determine how earnings are dispersed around their conditional mean. We use the National Child Development Study (NCDS) to examine this complementary aspect of the height--earnings relationship. Two comparisons organize the application. We first examine weekly-earnings dispersion in the pooled sample and within the male and female subsamples, using hourly pay and working hours to interpret the pooled pattern. We then hold the male hourly-earnings sample fixed and compare the restriction of constant conditional variance with the stronger restriction of an invariant conditional error distribution, tested by the fixed-scale procedure of \citet{EVK2008} (hereafter the EVK fixed-scale test). These comparisons distinguish variation associated with the earnings measure and sample composition from differences arising from the hypothesis being tested.

\subsection{Data and Implementation}\label{sec:ncds}

Much of the existing research on the height premium, including \citet{CasePaxson2008}, has focused on mean labor-market outcomes and the role of cognitive ability and occupational sorting. Our analysis instead examines conditional earnings dispersion. This distinction is relevant because weekly earnings reflect both hourly remuneration and the number of hours worked, whereas hourly earnings isolate remuneration per hour in the accounting identity used below. A height-related change in weekly-earnings dispersion may therefore have a different interpretation from a change in hourly-pay dispersion. We use the NCDS Age 33, Sweep 5, 1991 data collection, which follows the 1958 British birth cohort \citep{NCDS1991}. The complete-case sample contains 6,785 employed respondents, including 3,703 men and 3,082 women. The outcomes are log gross weekly earnings and log hourly earnings, with hourly earnings defined as weekly earnings divided by usual weekly hours. Adult height is recorded to the nearest centimeter, and all recorded ties are retained.

The primary analyses trim 1\% from each tail of the relevant earnings-level distribution. For weekly earnings, trimming is applied once to the pooled sample, and the sex-specific analyses use subsets of that retained sample. Male hourly earnings are trimmed within the male sample. The proposed test uses $a=C s_X n^{-1/3}$ with $C=4$ in the main tables, while $C=3.5$ and $C=4.5$ are reported as sensitivity checks in Section~C.2 of the supplementary material. All tests in this empirical application use $B=5{,}000$ bootstrap draws.

\subsection{Conditional Variance of Weekly Earnings}\label{sec:ncds_weekly}

Both CvM and KS reject constant conditional variance in pooled log weekly earnings at the minimum attainable bootstrap $p$-value of $0.0002$ (Table~\ref{tab:ncds_weekly}). In contrast, neither statistic rejects in the male or female subsample, and this contrast persists at the neighboring bandwidth coefficients reported in Section~C.2.1 of the supplementary material. These subsamples are formed after applying a single trimming rule to pooled weekly earnings, so their different test outcomes do not arise from separate sex-specific tail cutoffs. The evidence points to a pronounced difference between pooled and within-sex analyses, but it does not establish that either subgroup is homoskedastic or that pooling is the sole source of the pooled rejection.

\begin{table}[!htbp]
\centering\small
\caption{NCDS weekly earnings: tests of constant conditional variance}\label{tab:ncds_weekly}
\begin{tabular}{lrrr}
\toprule
Sample & $n$ & CvM & KS \\
\midrule
Pooled & 6,649 & 0.0002 & 0.0002 \\
Men within pooled-trimmed sample & 3,650 & 0.5939 & 0.6429 \\
Women within pooled-trimmed sample & 2,999 & 0.5387 & 0.4357 \\
\bottomrule
\end{tabular}
\begin{minipage}{0.94\linewidth}\footnotesize
\emph{Notes:} Entries are bootstrap $p$-values. All rows use the common pooled 1\%--99\% weekly-earnings trimming rule, $C=4$, and $B=5{,}000$. The minimum attainable $p$-value is $1/(B+1)\approx0.0002$.
\end{minipage}
\end{table}
\FloatBarrier

\begin{figure}[!htbp]
\centering
\includegraphics[width=0.92\textwidth]{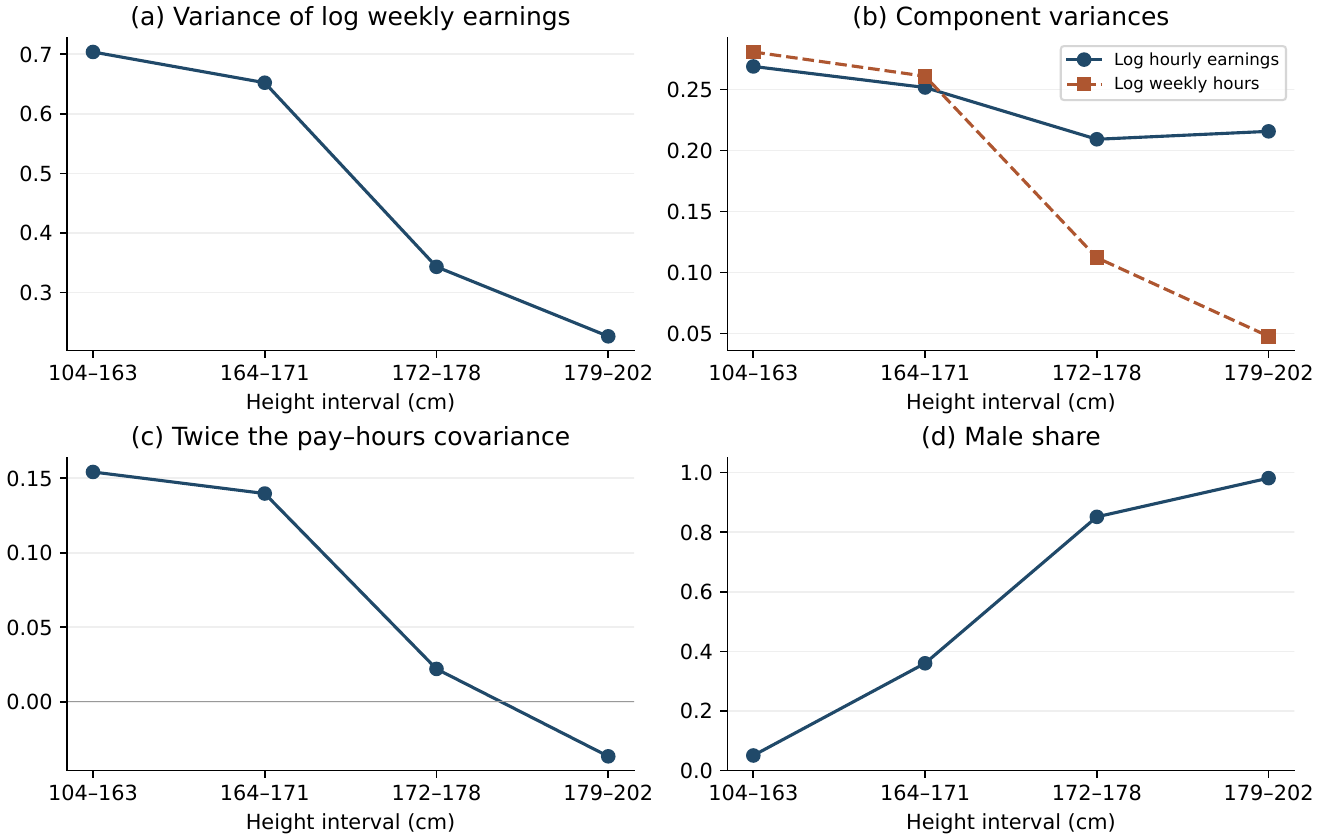}
\caption{Weekly-earnings dispersion, hours dispersion, pay--hours covariance, and sex composition across fixed height intervals. All panels use the primary pooled-weekly sample ($n=6{,}649$). Points are descriptive bin-level summaries; lines connect the summaries.}
\label{fig:ncds_weekly_decomp}
\end{figure}

The contrast has two economically distinct dimensions. The first concerns how weekly earnings combine pay rates with hours of work. Variation in hours and its association with hourly pay can change the dispersion of weekly earnings even when hourly-pay dispersion changes comparatively little. Let $W_w$, $W_h$, and $H$ denote weekly earnings, hourly earnings, and weekly hours. Since $\log W_w=\log W_h+\log H$,
\begin{align*}
\operatorname{Var}(\log W_w\mid X)
={}&\operatorname{Var}(\log W_h\mid X)
+\operatorname{Var}(\log H\mid X)\notag\\
&+2\operatorname{Cov}(\log W_h,\log H\mid X).
\end{align*}
Figure~\ref{fig:ncds_weekly_decomp} presents the components of this identity across four fixed height intervals in the pooled sample used for the weekly-earnings test. From the lowest to the highest interval, the variance of log weekly earnings declines from 0.704 to 0.227. The corresponding variance of log hourly earnings changes much less, from 0.269 to 0.216. In comparison, the variance of log weekly hours falls from 0.281 to 0.048, while twice the covariance between log hourly earnings and log weekly hours moves from 0.154 to $-0.037$. Thus, much of the observed difference in weekly-earnings dispersion between the extreme height intervals is accompanied by changes in the dispersion of hours and in the association between pay and hours, rather than by a comparably large change in hourly-pay dispersion. This accounting comparison helps explain why weekly and hourly earnings can display different dispersion patterns. The plotted variances are calculated within height intervals and may also reflect changes in conditional means inside those intervals. They are descriptive summaries, not pointwise estimates of conditional variance or evidence that working hours causally mediate an effect of height.

The second dimension concerns the composition of the pooled sample. The proportions of men and women vary across the height intervals, and the pooled variance combines within-group dispersion with differences in their conditional mean earnings. For the sex indicator $G$,
\begin{align*}
\operatorname{Var}(Y\mid X)
={}&\mathbb E\{\operatorname{Var}(Y\mid X,G)\mid X\}
+\operatorname{Var}\{\mathbb E(Y\mid X,G)\mid X\}.
\end{align*}
The first term averages sex-specific conditional variances, while the second reflects differences in sex-specific conditional means at a given height. Both terms can change as group proportions vary with height, even if the within-sex variance functions were constant. In this sense, the pooled rejection and the two subgroup nonrejections are compatible with a role for sample composition. The identity does not, however, allocate the observed rejection to a particular component, and subgroup nonrejection is not evidence that the corresponding variance function is exactly constant. Together with the weekly-pay decomposition, it provides an economic interpretation of the pooled result without attributing it to a single mechanism.

\FloatBarrier

\subsection{Conditional Variance versus an Invariant Error Distribution}\label{sec:ncds_hourly}

The weekly-earnings analysis changes the composition of the sample while retaining the same conditional-variance restriction. A complementary comparison is possible in the male hourly-earnings sample. Here we keep both the sample and the earnings measure fixed, but change the strength of the restriction imposed on the regression error.

\begin{samepage}
Male hourly earnings provide a direct comparison of the two restrictions. Writing $Y$ for log hourly earnings, the variance null is
\begin{equation*}
H_0^{\mathrm{var}}:\operatorname{Var}(Y\mid X)=\sigma^2,
\end{equation*}
whereas the EVK fixed-scale null is the homoskedastic version of the specification test of \citet{EVK2008},
\begin{equation*}
H_0^{\mathrm{ind}}:Y=m(X)+e,\qquad e\Perp X.
\end{equation*}
Under finite second moments, $H_0^{\mathrm{ind}}$ implies $H_0^{\mathrm{var}}$, since the former requires the conditional error distribution to be invariant with respect to $X$, whereas the latter restricts only its variance.
\end{samepage}

For the primary sample trimmed by 1\% at each tail, the proposed CvM and KS tests yield $p$-values of 0.6837 and 0.8760, respectively, and thus do not reject the null of constant conditional variance. For the same observations, the EVK fixed-scale test rejects the invariance of the conditional error distribution, with corresponding $p$-values of 0.0002 and 0.0032 (Table~\ref{tab:ncds_hourly}). These findings are compatible. The stronger restriction rules out height-related changes in the entire additive error distribution, whereas the variance null concerns only its second conditional moment. The comparison therefore illustrates how a rejection of error-distribution invariance does not by itself provide evidence of heteroskedasticity. Similar conclusions obtain under the neighboring 0.5\% and 2\% trimming rules. The untrimmed sample provides a qualification, as the EVK CvM statistic rejects the fixed-scale null, whereas the KS statistic does not. Nor should the variance-test nonrejection be interpreted as proof of homoskedasticity.

\begin{table}[!htbp]
\centering\small
\caption{Male hourly earnings: constant variance versus the EVK fixed-scale restriction}\label{tab:ncds_hourly}
\begin{tabular}{lrrr}
\toprule
Each-tail trimming & $n$ & Variance test CvM / KS & EVK fixed-scale CvM / KS \\
\midrule
None & 3,703 & 0.9680 / 0.9346 & 0.0210 / 0.0682 \\
0.5\% & 3,667 & 0.5971 / 0.8498 & 0.0006 / 0.0158 \\
1\% & 3,627 & 0.6837 / 0.8760 & 0.0002 / 0.0032 \\
2\% & 3,553 & 0.1174 / 0.0950 & 0.0002 / 0.0042 \\
\bottomrule
\end{tabular}
\begin{minipage}{0.94\linewidth}\footnotesize
\emph{Notes:} Entries are bootstrap $p$-values in CvM / KS order. The variance test uses $C=4$; the EVK fixed-scale test uses its cross-validated mean bandwidth and residual bootstrap. Both procedures use $B=5{,}000$. Trimming is applied to male hourly earnings before taking logarithms.
\end{minipage}
\end{table}
\FloatBarrier

\begin{figure}[!htbp]
\centering
\includegraphics[width=0.92\textwidth]{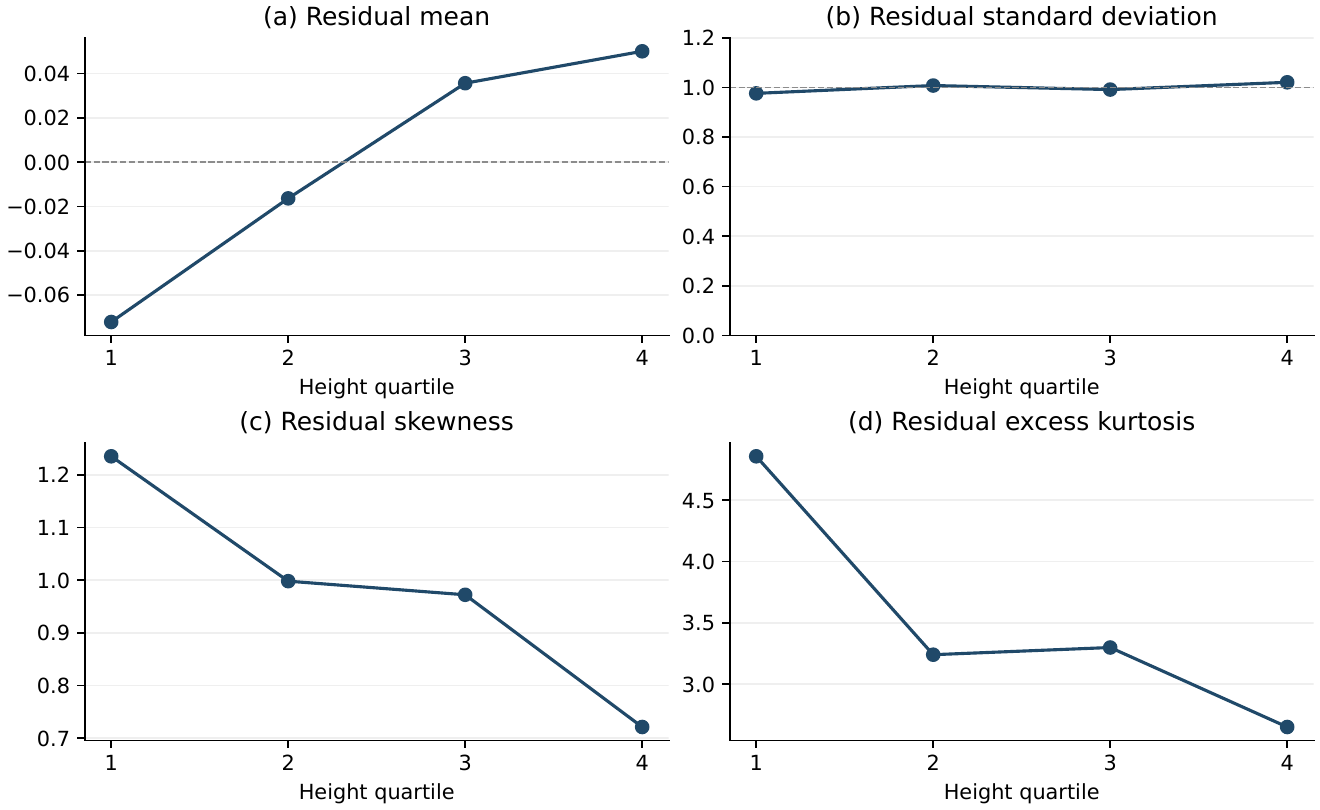}
\caption{Additive residual summaries across height quartiles for the primary male-hourly sample ($n=3{,}627$). Residuals use one global normalization and no height-dependent scale standardization. Points are descriptive summaries without inferential bands.}
\label{fig:ncds_hourly_resid}
\end{figure}

Figure~\ref{fig:ncds_hourly_resid} provides a descriptive view of how the residual distribution changes across height quartiles in the primary male-hourly sample. The additive residuals $\widehat u_i=Y_i-\widehat m_r(X_i)$, where $\widehat m_r$ is the cross-validated Nadaraya--Watson estimator defined in Section~C.3 of the supplementary material, are divided by one global residual root-mean-square, rather than by height-specific scale estimates. Their standard deviations remain in a narrow range, from 0.976 to 1.021. By contrast, residual skewness declines from 1.24 to 0.72 and excess kurtosis from 4.86 to 2.65. The coexistence of relatively stable residual dispersion and more pronounced differences in residual shape is consistent with the distinction between the two tested restrictions. These plots are not a separate inferential test of higher conditional moments, however, and they cannot establish which feature produces the EVK rejection. Differences in residual means across quartiles also leave a possible role for conditional-mean estimation.

\FloatBarrier

The NCDS evidence illustrates how the economic interpretation of earnings dispersion depends on the outcome, the sample composition, and the conditional restriction being tested.

\Needspace{10\baselineskip}
\section{Concluding Remarks}\label{sec:conclusion}

We develop direct tests of conditional homoskedasticity in nonparametric regression with random covariates, allowing higher-order conditional features of the error distribution to vary. The density-weighted third-order process admits a multiplier calibration for its nonpivotal null law. The theory establishes asymptotic validity under the null and consistency against fixed alternatives, and characterizes local behavior through a cumulative variance drift.

The experiments show that the proposed tests maintain rejection frequencies close to the nominal level under the null designs with varying higher-order conditional features. The power comparisons distinguish the sensitivity of CvM and KS to the shape of the variance departure, with cumulative cancellation helping to explain the weaker CvM power against the cosine alternative.

In the NCDS application, pooled weekly-earnings rejection calls for examining hours, pay--hours covariance, and group composition alongside the within-sex results. In the male-hourly sample, rejection of an invariant error distribution coexists with nonrejection of constant conditional variance. These findings illustrate the diagnostic value of testing the conditional second moment directly when heteroskedasticity is the object of interest.

\end{document}